\RequirePackage{amsmath}
\documentclass[10pt]{iopart}

\usepackage{graphicx} 
\usepackage{soul} 
\usepackage{float}
\usepackage{siunitx} 
\usepackage{amsmath}
\usepackage{hyperref}
\usepackage[T1]{fontenc}
\usepackage{lmodern}
\usepackage[table,xcdraw]{xcolor} 
\usepackage{pdflscape}
\definecolor{red_TRANSP}{rgb}{0.76,0.09,0.36}
\definecolor{green_LOCUST}{rgb}{0.298,0.686,0.314}
\definecolor{blue_ASCOT}{rgb}{0.132,0.5898,0.9531}

\begin{document}

\title[Fusion for high-value heat production]{Fusion for high-value heat production}

\author{S. H. Ward}
\address{Eindhoven University of Technology, Eindhoven, Netherlands}    
\author{M. Majeed}
\address{Eindhoven University of Technology, Eindhoven, Netherlands}
\author{N. J. Lopes Cardozo}
\address{Eindhoven University of Technology, Eindhoven, Netherlands}

\ead{s.h.ward@tue.nl}
\vspace{10pt}

\begin{abstract}

        Global consumption of heat is vast and difficult to decarbonise, but it could present an opportunity for commercial fusion energy technology.

        The economics of supplying heat with fusion energy are explored in context of a future decarbonised energy system. A simple, generalised model is used to estimate the impact of selling heat on profitability, and compare it to selling electricity, for a variety of fusion proposed power plant permutations described in literature.

        Heat production has the potential to significantly improve the financial performance of fusion over selling electricity. Upon entering a highly electrified energy system, fusion should aim to operate as a grid-scale heat pump, avoiding both electrical conversion and recirculation costs whilst exploiting firm demand for high-value heat. This strategy is relatively high-risk, high-reward, but options are identified for hedging these risks. We also identify and discuss new avenues for competition in this domain, which would not exist if fusion supplies electricity only.
    
\end{abstract}

%
%
%
%
\ioptwocol

\section{\label{sec:introduction}Introduction}

    Despite being the primary goal of most fusion research and development (R\&D) programmes \cite{FIA2022report}\cite{FIA2023report}\cite{muldrew2024conceptual}, commercial electrical power production is unlikely to be an optimal strategic outcome. Though highly uncertain, fusion electricity prices are typically estimated to be higher than wholesale electricity prices today \cite{lindley2023can}\cite{jo2021cost}\cite{entler2018approximation}\cite{hawker2020simplified}\cite{cook2002prospects}\cite{griffiths2022commercialisation}. Other studies have estimated that tokamaks with moderate capacity factors are almost unable to produce net electricity over their lifetime due to recirculating power requirements \cite{mulder2021plant}, or have their value forcibly limited \cite{takeda2016limitation} or destroyed \cite{takeda2020economic} by unplanned outages. Added to FOAK premia and little (or no \cite{grubler2010costs}) early-stage learning, these findings imply that fusion is unlikely to be competitive in electricity markets today. If fusion energy cannot compete, it cannot deliver any potential benefits to society.

    To increase the competitiveness of fusion, the value proposition of fusion energy must be improved \cite{nicholas2021re}. Whilst many traits influence and define this value proposition, it is the potential to generate financial and economic value that dictates investment decisions. Therefore, increasing the profitability of fusion devices is the key to enabling the technology to deliver safe, reliable, and carbon-free power at scale. The small but varied selection of studies that model fusion power in a wider energy system together mostly confirm the principle that costs play a pivotal role in how fusion penetrates electricity grids \cite{gi2020potential}\cite{bustreo2019fusion}\cite{schwartz2023value}. These studies also show the importance of context – the competition and external drivers of value.

    Besides cost, carbon intensity, and security of supply have become increasingly important drivers of economic value. As a result, significant policies are being implemented to incentivise a transition to a grid powered by an independent, low-carbon source of power. For example, the Inflation Reduction Act alone could deliver 46\% of decarbonisation targets in the United States \cite{bistline2023emissions}. Those technologies which stand to gain from these policies are already competitive in many places – specifically renewables and storage. Some estimate total internal system costs could already be reduced now by a transition to renewables, without any policy intervention \cite{way2022empirically}. A grid based around renewables is therefore highly likely in the future, but will behave fundamentally differently to the grid today due to renewables’ inherent intermittency. It is on this grid that fusion plants designed now should aim to be competitive and complimentary to alternative low-carbon technologies. This has been studied by Schwartz et al. \cite{schwartz2023value}, who showed that fusion penetration in deeply decarbonised grids depends on flexibility – a fundamentally different economic situation than experienced by nuclear plants providing baseload today. Similar conclusions have also been reached for advanced geothermal sources \cite{ricks2022value}. Handley et al. also examine how this changing context may create new opportunities for fusion to generate value, identifying early markets and cost targets \cite{handley2021potential}.

    An often ignored yet significant component of energy demand is heat: heat demand accounted for 50\% of total final energy consumption in 2018 \cite{iea2019analysis}, with industry consuming approximately half \cite{iea2022world}. Today’s heat supply is greatly dependent on fossil fuels (>75\%)\cite{iea2019analysis}. This means CO2 emissions from heat occupy 40\% of global CO2 emissions, with absolute emissions increasing 9\% between 2009-2018 \cite{iea2019analysis}. Decarbonising industry heat is much more challenging than grid electricity. Industry processes require stable sources of heat at specific temperatures, denoted low (<150°C), medium (150-400°C) to high (>400°C) temperatures \cite{iea2018cleanefficient}. With some studies showing \cite{madeddu2020co2} that a significant fraction of industry heat can now be electrified, the primary question is now cost. The shift from heat to electricity production will certainly introduce conversion costs that increase with process temperature. And even the costs of the underlying electricity too may be affected by the increasing demand \cite{iea2018cleanefficient} from industrial electrification, especially if growth outstrips capacity expansion – a plausible outcome if interconnection queues are large. Hence heat is a unique example of an opportunity – not threat – created by the energy transition for fusion.

    Few papers have discussed commercial fusion heat. Handley et al. \cite{handley2021potential} briefly cover process heat as a potential early market, deeming it challenging due to FOAK risk-intolerance, unattainable process temperatures (>700°C), and some processes not requiring external fuel sources. Very low-cost targets are also calculated for the indirect supply of high-temperature heat via hydrogen (LCOE < \$32/MWhr or < \$50/MWhr to compete with renewables or fossil gas with CCS respectively). Konishi et al. \cite{konishi2001use} is an early example that discusses the compatibility of fusion technology with various industrial processes. One specific example of fusion heat application is the GNOME fusion-biomass device \cite{ibano2013neutronics}\cite{ibano2011design}\cite{nam2020cost}, where a fusion device augments the ability for a biomass plant to create hydrogen and synthetic fuels or generate electricity. \cite{cano2022boosting} finds cogeneration of electricity and heat for district heating can be favorable for a specific tokamak model. To identify candidate heat use cases, Griffiths et al. \cite{griffiths2022commercialisation} sensibly turn to comparing with fission due to the lack of literature. \cite{hampe2016economic} quantified the options value of supplying both heat and electricity with fission plants, comparing it to the additional equipment costs. One key distinguishing advantage of fusion over fission heat is licensing and regulation. The European Union (EU) EUROPAIRS study specifically identified these as hurdles to supplying heat with fission \cite{angulo2012europairs}. A recent report \cite{royal2020nuclear} by the Royal Society identified SMRs as being the primary candidates for nuclear cogeneration due to their safety credentials, as well as their portability. However, \cite{vanatta2023technoeconomic} found that industry process heat from fission SMRs was competitive only if cogenerated electricity could be sold to the grids at times of low industry demand.

    Large questions remain regarding fusion for heat. Studies have hitherto focused on specific markets, time periods, and reactors – with specific designs, capabilities, and modes of operation. However, the design of commercial fusion devices remains uncertain, and the context they will operate in is rapidly changing. This necessitates a generalised analysis to bound the problem. Three questions should be answered:

    \begin{itemize}
        \item \textit{How much could generating heat change or improve the business case of individual fusion reactors?}
        \item \textit{What role could fusion heat play in a decarbonised energy economy?}
        \item \textit{Is supplying fusion heat technologically feasible?}
    \end{itemize}

    This paper addresses the first two questions by constructing an abstracted, quantitative model to describe heat produced by a fusion reactor, before discussing the qualitative implications of deploying this model at scale. Using this model, we determine the optimal electricity or heat production in the case where heat and electricity prices are stable, and how this might change the value of fusion energy. We then use this to discuss the possible role that fusion could play in reducing the costs of the wider energy system, as well as briefly discussing the implications of this role.

    \section{\label{sec:fusion_heat_generation}Fusion heat generation}

    Commercial fusion reactors are highly likely to generate primary energy as heat. This is because the most feasible candidate fuel mixtures (involving deuterium and tritium \cite{rider1995general}\cite{rider1997fundamental}) undergo nuclear reactions that release most of their energy via energetic neutrons. These neutrons are thermalised and captured in a “blanket” surrounding the reaction chamber. For D-T reactors, neutrons are designed to be captured by lithium, which is then transmuted new tritium fuel. A coolant – sometimes but not always the breeding material itself – is then pumped around and out of the blanket, before a working fluid – either the coolant or a secondary fluid via heat exchanger – is heated to generate electricity. A variety of blanket concepts exist, all in development and untested, each designed to operate with different breeders, coolants, thermodynamic cycles, and hence at different temperatures. A snapshot overview is given in Table \ref{tab:fusion_blankets}.

    \begin{table*}[!htbp]
        \centering

        \begin{tabular}{llll}
        \textbf{Blanket type}             & \textbf{Coolant type} & \textbf{Blanket outlet temperature {[}Celsius{]}} & \textbf{Reference}               \\
        HCPB (Li\textsubscript{4}SiO\textsubscript{4} + Be)   & He                & 500                     & \cite{cismondi2018progress}\cite{federici2019overview}      \\
        HCPB (Li\textsubscript{4}SiO\textsubscript{4} + Be)   & He + Water        & 500                     & \cite{tarallo2019preliminary}    \\
        HCPB (Li\textsubscript{4}SiO\textsubscript{4} + Li\textsubscript{2}TiO\textsubscript{3} + Be\textsubscript{12}Ti) & He                & 520                     & \cite{hernandez2019enhanced}     \\
        CCPB                              & CO\textsubscript{2}              & $\sim$500                                                 & \cite{wang2019first}      \\
        MLCB (solid Li + molten Pb)       & He or water       & 520                                                 & \cite{zhou2019progress}\cite{hernandez2019enhanced}     \\
        WCLL                              & LiPb + water         & 328                                             & \cite{edemetti2020optimization}\cite{cismondi2018progress}  \\
        HCLL                              & LiPb + He             & 500                                                 & \cite{cismondi2018progress}          \\
        DCLL                              & LiPb + He              & 548 for Pb, 445 for He                              & \cite{cismondi2018progress}      \\
        Aries-I                           & He                    & 900                                                 & \cite{najmabadi2006aries}        \\
        Aries-ST                          & LiPb + He              & 700                                                 & \cite{najmabadi2006aries}        \\
        Aries-AT                          & LiPb + He              & 1100                                                & \cite{najmabadi2006aries}        \\
        GAMBL                             & LiPb + He              & \textgreater{}1000                                  & \cite{tillack2022gambl}          \\
        SCYLLA                            & LiPb                  & $\sim$1000                                          & \cite{pearson2022overview}       \\
        Molten Salts                      &                       & At least 500                                        & \cite{wang2019first}             \\
                                            &                       & 500-600                                             & \cite{forsberg2020fusion}\cite{boullon2021molten}        \\
            \label{tab:fusion_blankets}
        \end{tabular}
        \caption{Table of selected fusion blanket technologies and characteristics related to heat output. Despite coolant technologies varying, it is typically the structural material which limits outlet temperature - for example 500C in the case of the EUROfer. Additionally, concepts essentially span the combinatorial space of neutron multiplier (lead or beryllium), breeding material phase (solid or liquid breeder), and coolant element (water, C02, helium, self-cooling). The most advanced blanket designs peak at approximately 1000C.}
    \end{table*}

    As described in \cite{cismondi2018progress}, the outlet temperatures of these blankets are typically limited by the choice and configuration of structural materials, which have strict operating windows to avoid damage. Electricity production requires conversion from heat using generators via thermodynamic cycles. Many decades of experience have made these cycles highly efficient (63\% as reported at the Chubu Electric Nishi-Nagoya power plant in 2018), but performance is still bounded by thermodynamics and thus the temperature difference between the coolant and the exhaust via Carnot’s theorem. This has motivated efforts to maximise blanket operating temperatures. After conversion to electricity, a significant fraction must then be recirculated, back to power reactor and plant systems. The composition and relative power requirements of these systems vary across fusion design concepts. Magnetic fusion devices specifically require power to be recirculated to magnet, cryogenic, and current drive systems, for example. However, regardless of the reactor design concept, controlled fusion ultimately requires systems to generate and extract high power densities – amounting to heating and cooling systems that inherently increase the recirculation fraction. For example, plasma heating and coolant pumping play significant roles in magnetic systems, and inertial fusion devices require significant power to be re-invested back into the driver. For pulsed devices like tokamaks and ICF facilities, the plant electricity consumption can be split into continuous and pulsed loads, the latter being required during or immediately before/after pulses (the ramping up/down period in tokamaks and the charging period in ICF devices). A continuous baseload electricity consumption is also required throughout ramping, pulsing, and periods in between – known as dwell periods. A high recirculating power fraction implies less electrical output available for generating revenues, so it must be minimised - or even avoided, if possible. The impact on overall efficiency can be significant; in combination with plant availability, the recirculating power fraction, which may typically reach 50\%, could mean some plants produce net-zero power \cite{mulder2021plant}.

\section{\label{sec:fusion_heat_model}Fusion heat model}

    \subsection{\label{subsec:revenues}Revenues}

\begin{figure*}[!htbp]
            \centering
                \includegraphics[width=0.7\textwidth]{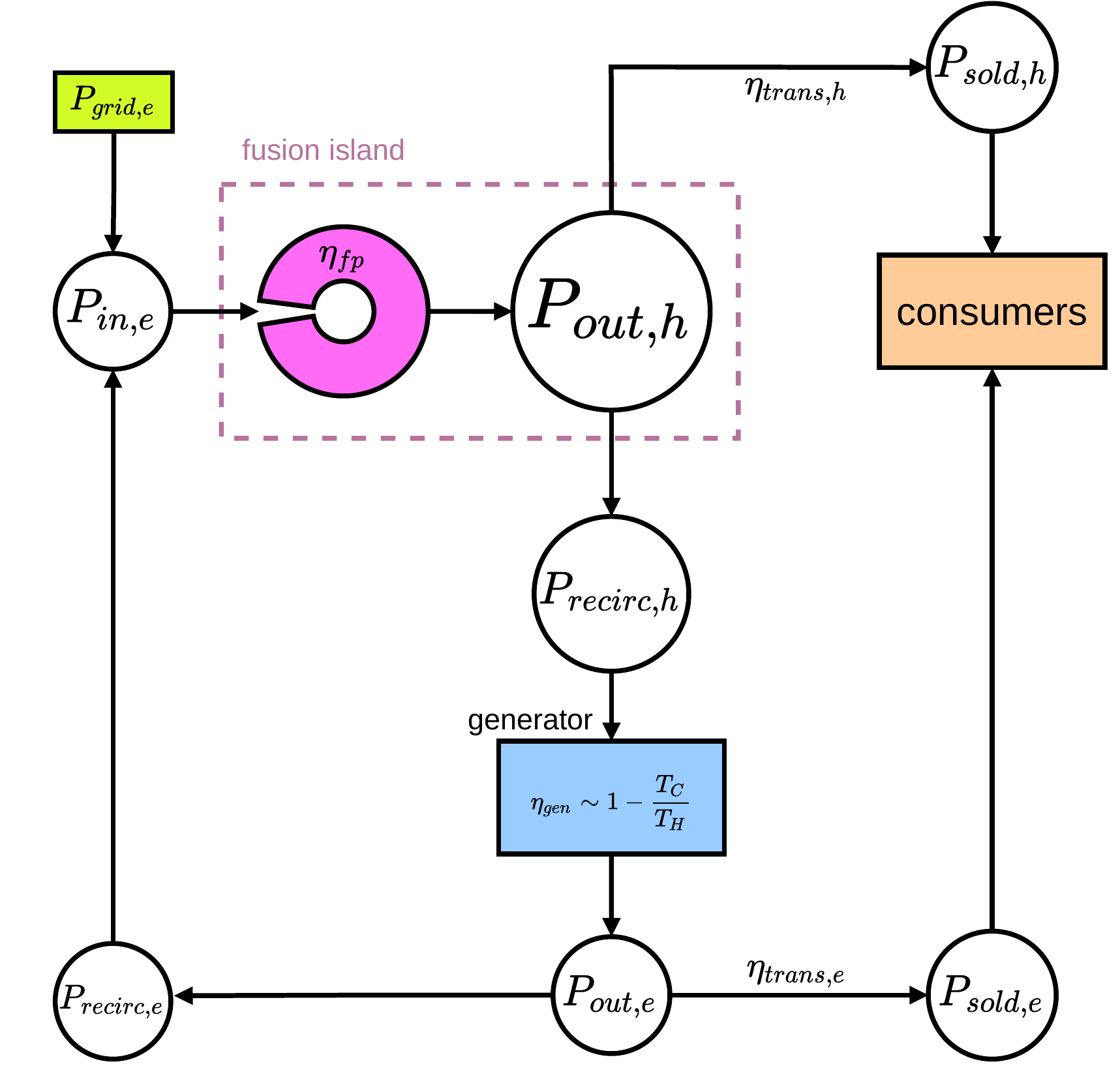}
                \caption{Schematic for power flows around and in/out of a simplified fusion cogeneration plant. The fusion reactor island itself is demarked by pink dash lines, including support systems like tritium and cryogenic plants, as well as heating and control systems, and efficiency losses. $\eta_{\mathrm{fp}}$ represents the multiplication of gross input power by fusion and subsequent nuclear reactions. A full breakdown is provided in text.}
                \label{fig:fusion_heat_pump}
        \end{figure*}

        An abstract model of a fusion plant for cogeneration is illustrated in Figure \ref{fig:fusion_heat_pump}. The equivalent model is also represented in equations \ref{eq:power_conservation_one} to \ref{eq:power_conservation_five}.

        \begin{eqnarray}
            P_{\mathrm{out,h}} = \eta_{\mathrm{fp}}P_{\mathrm{in,e}}
            \label{eq:power_conservation_one}
        \end{eqnarray}
        \begin{eqnarray}
            P_{\mathrm{in,e}} = P_{\mathrm{grid,e}} + P_{\mathrm{recirc,e}}
            \label{eq:power_conservation_two}
        \end{eqnarray}
        \begin{eqnarray}
            P_{\mathrm{out,e}} = P_{\mathrm{recirc,e}} + P_{\mathrm{sold,e}}
            \label{eq:power_conservation_three}
        \end{eqnarray}
        \begin{eqnarray}
            P_{\mathrm{out,h}} = P_{\mathrm{recirc,h}} + P_{\mathrm{sold,h}}
            \label{eq:power_conservation_four}
        \end{eqnarray}
        \begin{eqnarray}
            P_{\mathrm{out,e}} = \eta_{\mathrm{gen}}P_{\mathrm{recirc,h}}
            \label{eq:power_conservation_five}
        \end{eqnarray}

        The relevant sections of the fusion plant and its surroundings are represented as a directed network, with edges denoting flows of power. Nodes signify sources, sinks, junctions, or amplifiers. A description of the network components is now given.
        \\\\$P_{\mathrm{grid,e}}$
        \\    The electrical power supplied by the grid as input.
        \\\\$P_{\mathrm{in,e}}$
        \\    Gross electrical power input to the whole plant from both continuous and pulsed power draws averaged over a pulse. Since it is likely that thermal storage would benefit reactor economics, it can be argued that 100\% of the power required by the plant can be supplied by the plant during nominal operation. During maintenance periods or unplanned outages, power would be required from the grid, another fusion reactor, or long-duration storage, such as back-up generators. These periods are not modelled but their effects are captured in the average variable costs later.
        \\\\$\eta_{\mathrm{fp}}$
        \\    The power multiplication factor $P_{\mathrm{out,h}}$/$P_{\mathrm{in,e}}$. It measures how much primary power is released by a fusion plant for every unit of gross power supplied as input to the site. The amplification mechanism is the release of energy from fusion fuel and any secondary exothermic nuclear reactions, for example decay heat and certain neutron capture reactions, as well as some heat that conducts into the blanket – for example from divertor, radiation or first wall particle fluxes. As it measures gross power input, it includes any losses from powering overall plant systems and power losses during recirculation, for example dissipation during coolant pumping or electrical inefficiencies of plasma heating systems. To produce net power, a system with a non-zero recirculation must have $\eta_{\mathrm{fp}}$ > 1 / $\eta_{\mathrm{gen}}$.
        \\\\$P_{\mathrm{out,h}}$
        \\    The net power output from the blanket – that is, available to sell or be converted to electricity. In practice, $P_{\mathrm{out,h}}$ may differ between heat and electricity plants due to differences in design.
        \\\\$P_{\mathrm{sold, h}}$
        \\    Net heat power sold to consumers.
        \\\\$P_{\mathrm{recirc,h}}$
        \\    The gross amount of heat power directed back to the plant for electrical conversion. Transmission losses here are assumed to be negligible or contribute an additional error term to the conversion efficiency.
        \\\\$\eta_{\mathrm{gen}}$
        \\    The average electrical conversion efficiency.
        \\\\$P_{\mathrm{out,e}}$
        \\    The gross output of electrical conversion equipment.
        \\\\$P_{\mathrm{recirc,e}}$
        \\    The amount of electrical power recirculated for powering overall plant systems.
        \\\\$P_{\mathrm{sold,e}}$
        \\    The net electrical power sold to consumers.
        \\\\$\eta_{\mathrm{trans}}$
            The transmission efficiencies of heat and electricity. For simplicity of comparison, transmission efficiency is assumed to be perfect for all outputs. Whilst electricity transmission characteristics are well known, the same is not true for heat.

        In addition to the model equations, switches can be defined to describe a general device which can shift input and output between grid/recirculation and heat/electricity respectively:

        \begin{eqnarray}
            \xi \equiv \frac{P_{\mathrm{grid,e}}}{P_{\mathrm{in,e}}} = 1 - \frac{P_{\mathrm{recirc,e}}}{P_{\mathrm{in,e}}}
            \label{eq:definition_xi}
        \end{eqnarray}

        \begin{eqnarray}
            \epsilon \equiv 
            \frac{P_{\mathrm{sold,h}}}{P_{\mathrm{sellable,h}}}
            \label{eq:definition_epsilon}
        \end{eqnarray}

        where

        \begin{eqnarray}
            P_{\mathrm{sellable,h}} \equiv P_{\mathrm{out,h}} - \frac{P_{\mathrm{recirc,e}}}{\eta_{\mathrm{gen}}}
            \label{eq:definition_power_sellable}
        \end{eqnarray}

        and where both $\xi$ and $\epsilon$ can independently vary between 0 and 1.

        The objective of firms is to maximise profit, which is driven by revenues from selling power. The net revenues from selling power are defined as:

        \begin{eqnarray}
            R_{\mathrm{power}}\equiv 
            R_{\mathrm{h}}+
            R_{\mathrm{e}}-
            C_{\mathrm{grid}}
            \label{eq:definition_revenue_net}
        \end{eqnarray}

        where

        \begin{eqnarray}
            R_{\mathrm{e}}=
            P_{\mathrm{sold,h}}
            M_{\mathrm{e}}
            \eta_{\mathrm{trans,e}}
            Cp_{\mathrm{e}}
            \label{eq:definition_gross_revenue_electricity}
        \end{eqnarray}

        \begin{eqnarray}
            R_{\mathrm{h}}=
            P_{\mathrm{sold,h}}
            M_{\mathrm{h}}
            \eta_{\mathrm{trans,h}}
            Cp_{\mathrm{h}}
            \label{eq:definition_gross_revenue_heat}
        \end{eqnarray}

        \begin{eqnarray}
            C_{\mathrm{grid}}=
            P_{\mathrm{grid,e}}
            M_{\mathrm{e}}
            \eta_{\mathrm{trans,h}}
            Cp\left (Cp_{\mathrm{h}},Cp_{\mathrm{e}} \right )
            \label{eq:definition_gross_cost_grid}
        \end{eqnarray}

        \noindent and $M$ represents the market price of each commodity and $\eta_{\mathrm{trans}}$ gives the transmission efficiency. Here, $Cp_{\mathrm{h}}$ and $Cp_{\mathrm{e}}$ are defined as the capacity factor during heat or electricity production respectively, with $Cp$ the capacity factor of producing power at any time regardless of type (in reality higher than $Cp_{\mathrm{h}}$ or $Cp_{\mathrm{e}}$ individually due to the option to switch outputs in the case of unplanned outages in one channel). For the sake of simplicity, both $\eta$ and $Cp$ are assumed to be 1. Only operating commercial fusion plants will reduce the large uncertainty on these parameters – other than $eta_{\mathrm{trans,e}}$, which is accurately known today. Indeed it has been shown that the true impact of availability, especially on a future grid, is highly context-dependent \cite{schwartz2023valuing}. Nevertheless, the average effects of these parameters could be captured as additional variable costs.

        Expressing the net revenue as a fraction of the overnight capital cost (OCC) gives:

        \begin{equation}
        \begin{aligned}
        \frac{R_{\mathrm{power}}}{C_{\mathrm{OCC}}} & = 
        \tau_{\mathrm{h}}^{-1}
        [
            \frac{M_{\mathrm{e}}}{M_{\mathrm{h}}} \left( \eta_{\mathrm{gen}}-\frac{1}{\eta_{\mathrm{fp}}}\right) 
            \\
            & +
            \epsilon \left( 
                1+\frac{\xi-1}{\eta_{\mathrm{gen}}\eta_{\mathrm{fp}}}
            \right) 
            \left( 
                1-\eta_{\mathrm{gen}}\frac{M_{\mathrm{e}}}{M_{\mathrm{h}}}
            \right)
        ]
        \end{aligned}
        \label{eq:gross_revenue_normalised_simple}
        \end{equation}

        Here it is assumed that the OCC remains constant over all permutations of $\xi$ and $\epsilon$. In reality, an inflexible plant will have lower OCCs than the flexible plant, due the lack of conversion infrastructure. $\tau_{\mathrm{h}}$ is given in equation \ref{eq:definition_capacity_payback_time} and denotes the capacity payback time: the minimum possible time for each installed watt to repay its overnight cost. It is the nameplate capacity cost divided by the commodity price:

        \begin{eqnarray}
            \tau_{\mathrm{h}} \equiv \frac{\mathrm{OCC}}{P_{\mathrm{out,h}} M_{\mathrm{h}}} 
            \label{eq:definition_capacity_payback_time}
        \end{eqnarray} 

        By inspecting the roots of expression \ref{eq:gross_revenue_normalised_simple}, one can discern the optimal strategy for firms: if the energy price ratio exceeds the electrical conversion efficiency, then output heat \textit{and} take electricity from the grid, otherwise simply output electricity. Importantly, without switching costs and sufficient demand, there is nothing to be gained from producing a mixture of outputs. This can be explained with value: the goal is always to convert low-value inputs to high-value outputs, such that the increase in value must outstrip any reductions in output volume (in the case of conversion losses). This decision criterion is summarised in the statement \ref{eq:cases_operating_modes}:

        \begin{equation}
            \frac{M_{\mathrm{h}}}{M_{\mathrm{e}}} 
            \left\{
            \begin{array}{ll}
                >\eta_{\mathrm{gen}} & \text{max }\epsilon \text{, max } \xi \\
                <\eta_{\mathrm{gen}} & \text{min }\epsilon 
            \end{array}
            \right.
            \label{eq:cases_operating_modes}
        \end{equation}

\begin{figure}[!htbp]
            \centering
                \includegraphics[width=0.3\textwidth]{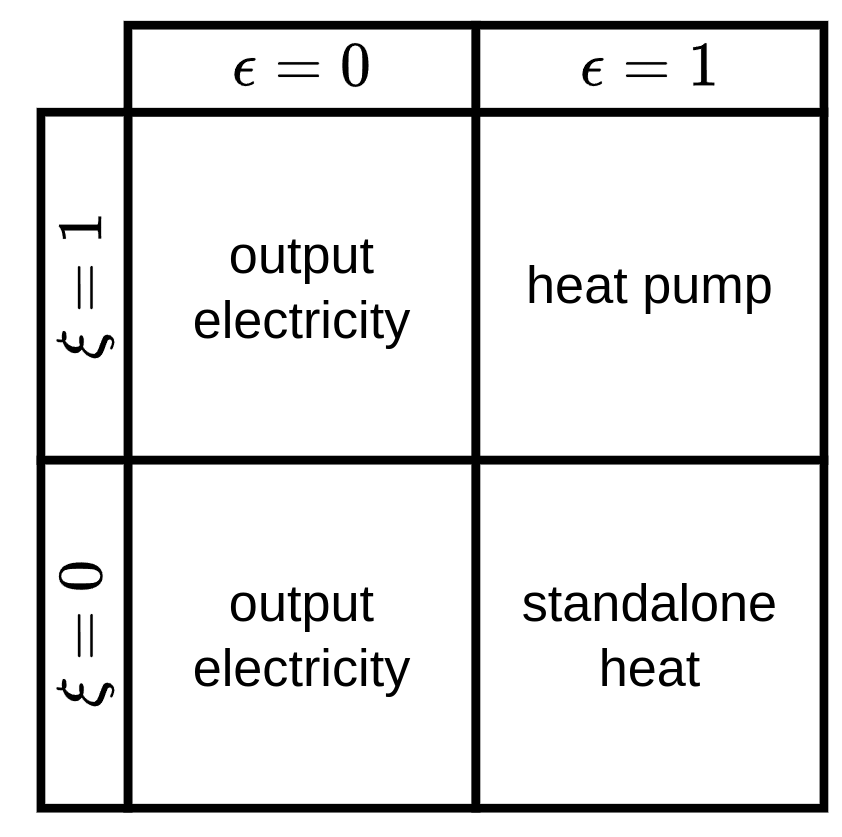}
                \caption{Decision matrix linking values of free parameters in equation \ref{eq:gross_revenue_normalised_simple} to their corresponding modes of operation. $\xi$ controls inputs and $\epsilon$ controls outputs.}
                \label{fig:fusion_cogeneration_permutations}
        \end{figure}

        These resulting operational decision rules are shown in Figure \ref{fig:fusion_cogeneration_permutations}.

    \subsection{\label{subsec:value}Value}

        To quantitatively compare fusion heat and electricity from the perspective of investors, a simple model to estimate net present value (NPV) is built. NPV is defined as 
        
        \begin{eqnarray}
            \mathrm{NPV} \equiv \sum_{t = 0}^{\infty} \frac{\mathrm{CF}_{t}}{(1+r)^{t}}
            \label{eq:definition_NPV}
        \end{eqnarray}
        
        where $CF$ denotes the net discounted cash flows (DCF) in each year $t$ and $r$ represents a discount rate. NPVs are designed to measure the financial value of an asset today by using the volume, timing, and likelihood of cash flows. For a fusion plant project, the cash flows in year $t$ are likely to be made up of 

        \begin{eqnarray}
            \begin{aligned}
                \mathrm{CF}_{t} & = 
                (R_{\mathrm{power,t}} - \mathrm{Vc}_{t} - \mathrm{In}_{t} - \mathrm{Dp}_{t})(1-T)
                \\ & + \mathrm{Dp}_{t} - \mathrm{Pr}_{t}
            \label{eq:cash_flows}
            \end{aligned}
        \end{eqnarray}

        where $T$ is the tax rate, Vc the averaged variable costs (all OPEX), Dp is depreciation, Pr the principal repayments and In the interest. Whilst DCFs typically do not make assumptions about the project capital structure (see Adjusted Present Value or post-tax weighted-average cost of capital), we do so here to to estimate the return to equity holders. The plant is assumed to be financed entirely with debt (debt:equity ratio of 1), and a simple debt structure is included in the DCF model - such that the discount rate essentially captures the cost of equity. We assume debt covers all expenditure prior to the end of the first year of operations – materials, labour, commissioning, etc. Debt repayments are assumed to be fixed, annual annuities over $\tau_{\mathrm{debt}}$ years, starting from the initial year of operation. Interest and principal repayments are separated to account for the tax shield. Meanwhile, assets are depreciated on a straight-line basis over the plant lifetime, meaning Dp is a constant. No salvage value is assumed, however this could become significant if recyclable, low-waste reactors are developed. Vc captures the average annual variable operating costs over the plant lifetime, including reactor fueling and peripheral plant systems, such as tritium plants. It could also account for the value lost to outages – unplanned and planned – not factored in dynamically to the power flow model. However, as the frequency and duration of these events are highly uncertain, they are not considered here – meaning VC represents a lower bound on variable costs.

        A compounding discount is applied to future cash flows given by the discount rate $r$. If the nominal cash flows above are near-constant over the plant lifetime – the exception being interest and principal payments in the presence of a tax – then a closed-form expression for the the average present value can be calculated, without numerically evaluating the sum. To find the average present value of a series of cashflows, one can define:
        
        \begin{eqnarray}
            \sum^{N}_{t=0} \frac{CF}{(1+r)^{t}} = N\times \kappa \times CF
            \label{eq:definition_average_discount}
        \end{eqnarray} 

        where

        \begin{eqnarray}
            \kappa \equiv \frac{1}{N}\frac{1-(1+r)^{-N}}{1-(1+r)^{-1}} = \kappa(N).
            \label{eq:expression_kappa}
        \end{eqnarray} 
        
        If interest also applies to the cash flow, one can also define

        \begin{eqnarray}
            \chi \equiv \frac{1}{N}\frac{1-\frac{1+i}{1+r}^{N}}{1-\frac{1+i}{1+r}}
            \label{eq:expression_chi}
        \end{eqnarray} 

        where $i$ is the interest rate. In many cases, the discount and interest rates are set equal, however here we allow for larger discount rates to simulate exogenous sources of risk - for example debt and counterparty risk. All cashflows are averaged over the project lifetime, as measured from the start of operations until the final day of operations. Likewise, cashflows are modelled as occurring at the end of the year, meaning an extra period of discounting is applied to all cashflows. The construction time is accounted for by appling a factor $(1+r)^{-\tau_{\mathrm{build}}}$ to all cashflows. Interest during construction is similarly included. Finally, decommissioning costs are ignored here; they are uncertain and contribute little to the NPV calculation due to their heavy discounting and purported low costs for fusion.




        Table \ref{tab:model_typical_parameters} shows values from literature for the parameters above. It must also be mentioned that values quoted in literature do not account for all the efficiency terms included in the parameters herein; these values are optimistic. However, to compare relative performance of heat and electricity plants, these values form a sensible first-order estimate. The return on capacity is given by the electricity price [\$/kWhr] * 8960 / (1000 * capacity cost [\$/W]). As a figure of merit, for an electricity price of \$0.1/kWhr and capacity cost of \$2/W, the return on capacity is 0.45. 

        \begin{table*}[]
        \begin{tabular}{lllll}
            \textbf{Parameter}                    & \textbf{Range {[}literature{]}} & \textbf{Range used}  & \textbf{Baseline} & \textbf{Reference}                                                                                                                                                                                                                                                                                                                                                                          \\
            $\tau_{\mathrm{build}}$ {[}yr{]}      & 10                                                     & 5-15                                       & 10                & \cite{entler2018approximation}                                                                                                                                                                                                                                                                                                                                             \\
            $\tau_{\mathrm{debt}}$ {[}yr{]}       & 24-27                                                  & 10-30                                      & 25                & \cite{EconomicsNuclearEnergy}\cite{FinancingNuclearEnergy}                                                                                                                                                                                                                                                                                                \\
            Project operating lifetime {[}yr{]}   & 40                                                     & 20-50                                      & 40                & \cite{entler2018approximation}\cite{takeda2020economic}\cite{syblik2023techno}                                                                                                                                                                                                                                                           \\
            Return on capacity {[}/yr{]}          & 0.1-1.5                                                & 0.1-1.5                                    & 0.5               & \begin{tabular}[c]{@{}l@{}}\cite{entler2018approximation}\cite{takeda2020economic}\cite{woodruff2017conceptual}\cite{moir1992hylife}\cite{hawker2020simplified}\cite{han2009revised}\end{tabular}                                                                                  \\
            $\eta_{\mathrm{fp}}$                  & 4.3-17                                                 & 4-20                                       & 15                & \cite{coleman2019blueprint}\cite{kerekevs2023operational}\cite{kerekevs2023operational}\cite{entler2018approximation}\cite{Krotez2023conceptual}\cite{moir1992hylife}                                                                                                                 \\
            $\eta_{\mathrm{gen}}$                 & 0.2-0.6                                                & 0.2-0.6                                    & 0.4               & \cite{woodruff2017conceptual}\cite{Krotez2023conceptual}\cite{moir1992hylife}\cite{hawker2020simplified}\cite{entler2018approximation}\cite{warmer2016system}\cite{cano2022boosting}\cite{syblik2023techno}\cite{syblik2023techno} \\
            OPEX (/OCC) {[}/yr{]}                 & 0.01-0.08                                              & 0.01-0.1                                   & 0.05              & \cite{nam2020cost}\cite{takeda2020economic}\cite{woodruff2017conceptual}\cite{hawker2020simplified}\cite{han2009revised}\cite{turnbull2015investigating}                                                                                                                              \\
            Conversion CAPEX (/OCC )              & 0-0.12                                                 & 0-0.1                                      & 0                 & \cite{woodruff2017conceptual}\cite{EconomicsNuclearEnergy}\cite{FinancingNuclearEnergy}\cite{syblik2023techno}                                                                                                                                                                                                          \\
            Storage CAPEX (/OCC)                  & 0-0.05                                                 & 0-0.1                                      & 0                 & \cite{takeda2020economic}                                                                                                                                                                                                                                                                                                                                                  \\
            Tax rate {[}\%{]}                     & 30                                                     & 0-30                                       & 30                & \cite{entler2018approximation}                                                                                                                                                                                                                                                                                                                                             \\
            Cost of capital {[}\%{]}              & n/a                                                    & $r$/2 - $r$                                & 5                 &                                                                                                                                                                                                                                                                                                                                                                                             \\
            Discount rate $r$ {[}\%{]}            & free parameter                                         & 3-12                                       & 6                 & \cite{entler2018approximation}\cite{takeda2020economic}\cite{syblik2023techno}                                                                                                                                                                                                                                                           \\
        \end{tabular}
            \caption{Table of model parameters and related quantities, as well as the ranges extracted from literature. The final baseline and value range used in the DCF Monte Carlo analysis is also shown. Variation in fusion concept yields little difference in the average values, except in a discount required for storage CAPEX and conversion CAPEX. Where required, prices were converted to 2024 USD using a CPI inflation calculator \cite{CPI_inflation}. Capacity payback time was calculated from capacity cost and assuming a relatively low wholesale energy price of 0.1 USD(2024)/kWhr. In studies in which decay heat was not included, only fusion power was counted to form a lower estimate.}
            \label{tab:model_typical_parameters}
        \end{table*}

    \section{\label{sec:fusion_heat_performance}Model results: The potential increase in profitability from supplying heat}

        The results from the model are now plotted to illustrate the impact of selling heat on plant revenues and profitability. Capacity factor is assumed to be 100\% for the sake of simplicity of comparison, whereas in reality it is likely to be significantly lower for FOAK plants. Using the baseline values given in Table \ref{tab:model_typical_parameters}, Table \ref{tab:baseline_NPV_YoY_return} shows the lifetime NPV (in units of overnight capital cost) and average annual NPV growth generated by a plant operating solely in each of the four modes above over its lifetime. Though NPV is positive in all cases, supplying heat leads to a roughly 30x increase in NPV. Switching electricity input sources also increases NPV by a further 20\%.

        \begin{table}[]
            \begin{tabular}{l|ll}
                NPV (real growth) & $\epsilon=0$            & $\epsilon=1$ \\ \hline
                $\xi=0$                   & 0.05 (0.12\%)   & 1.5 (2.34\%) \\
                $\xi=1$                   & 0.05 (0.12\%)   & 1.8 (2.62\%)
            \end{tabular}
            \caption{NPV generated by baseline fusion plants operating in each of the four modes described - toggling heat vs electricity output (columns) and grid vs recirculating electricity input (rows). NPV is expressed in units of overnight capital cost, whilst the number in brackets denotes the annual NPV growth during operation. Switching to heat output increases revenues by 30x whilst switching input sources may add a further 20\%.}
            \label{tab:baseline_NPV_YoY_return}
        \end{table}

\begin{figure*}[!htbp]
            \centering
                \includegraphics[width=0.7\textwidth]{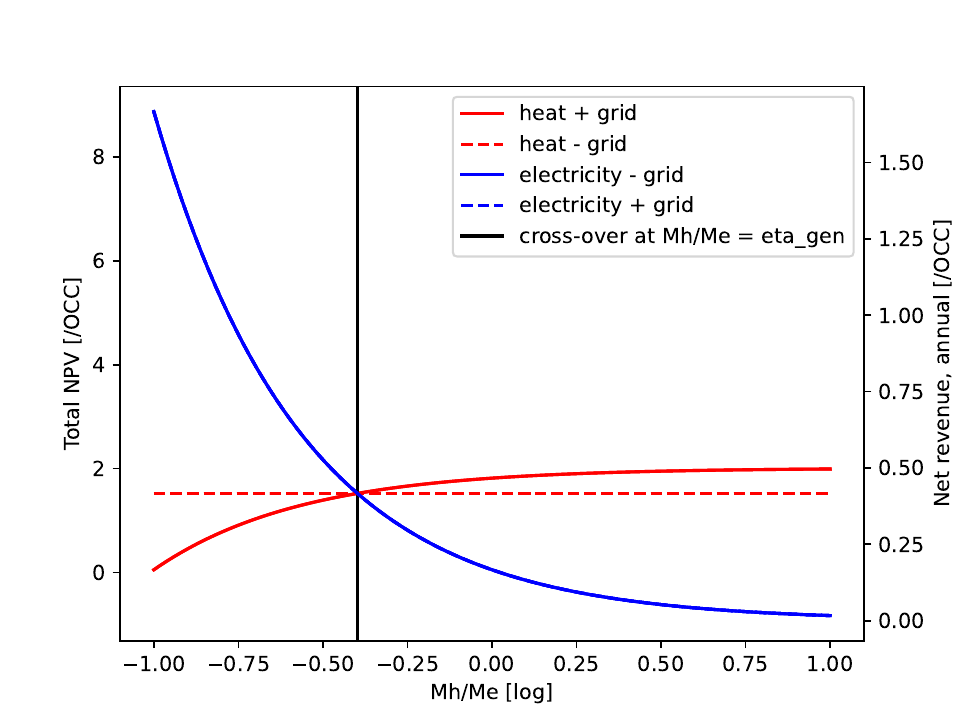}
                \caption{Revenues (right axis) generated for each of the operating modes from Figure \ref{fig:fusion_cogeneration_permutations}, as well as the corresponding NPV (left axis) for a plant with baseline characteristics given in Table \ref{tab:model_typical_parameters}. Also shown is the point at which revenues and net profit are equal, where the loss in output volume when converting to electricity is balanced by the increase in relative value. Quantities are plotted against the ratio of heat to electricity prices, and the latter is varied. This means the revenues for the standalone heat plant are constant, but would take a similar form as electricity revenues if instead only the heat price was varied.}
                \label{fig:net_revenue_NPV_price_ratio}
        \end{figure*}

        To quantify the trade-offs of operating in each mode under uncertain market conditions, the net revenues (after accounting for electricity input costs) for the baseline fusion plant are shown for different heat and electricity price ratios in Figure \ref{fig:net_revenue_NPV_price_ratio}. Price symmetry is assumed between input and output electricity. The return on capacity is held constant, implying electricity price is varied. Alternatively, the effects of varying the heat price can be understood by realising that heat revenues have a simple linear dependence on heat price, and that electricity revenues remain unaffected. However, the purpose here is to illustrate the relative trade-off between selling heat and electricity about a given absolute price point – rescaling that point does not change the overall behavior. In reality, there is volatility due to exposure to markets: to markets for electricity supply and capacity when buying or selling electricity, or to markets for solid fuel in the case of generating heat today. In this sense, the results here could also be interpreted as the case where heat prices are relatively inviolate over the plant lifetime compared with electricity prices. For this comparison, the conversion and storage CAPEX were set to zero. At equal electricity and heat prices, only electrical conversion losses distinguish the revenues generated in each operating mode, with revenues from fusion heat pumps greater by a factor $1/\eta_{\mathrm{gen}}$. Revenues are therefore balanced amongst all operating modes at the point where electricity prices are rescaled by a factor $1/\eta_{\mathrm{gen}}$, which is shown in black.

        If prices vary, the shift in relative value between heat and electricity introduces another discerning factor for the revenues generated by each operating mode. Electricity revenues have a linear dependence on electricity prices (the reciprocal plotted here), never reaching zero. This option has the potential to yield the largest returns over a small domain where electricity prices are sufficiently high. Selling heat without dependence on the grid understandably has no dependence on electricity prices. As such, there is no exposure to volatile grid prices and little risk. However, this option never produces the highest revenues, making the reward relatively small. Finally, converting grid electricity to heat has the highest potential upside across most of the domain. However, this mode is unique in that revenues can go negative – signifying input costs outweigh output value. This effect diminishes as $\eta_{\mathrm{fp}} \rightarrow \infty$, when the required input electricity nears zero and this mode resembles the case where heat is sold without grid input. Otherwise, costs could be incurred if heat offtakers cannot undertake demand response or if there are significant start-up or shutdown costs. These scenarios constitute three separate risk-reward spreads, which are layed out in Figure \ref{mat:risk_reward_spread} below.

        \begin{figure}[!htbp]
            \centering
                \includegraphics[width=0.3\textwidth]{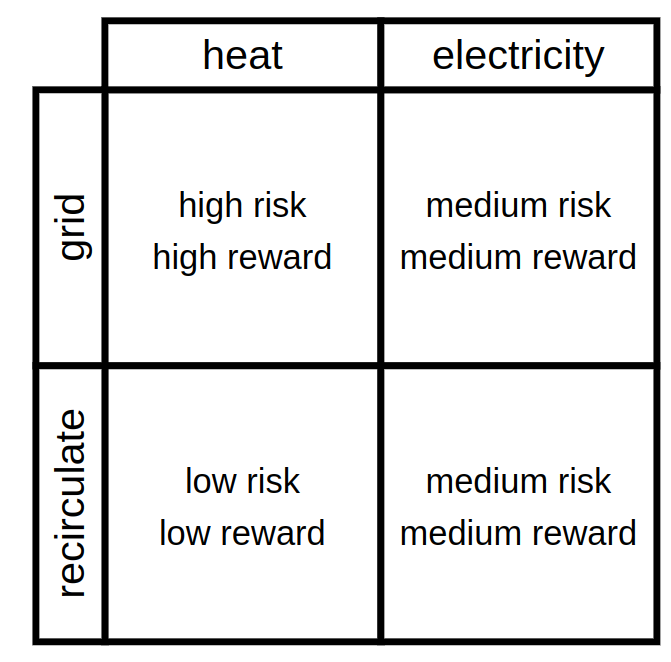}
                \caption{Decision matrix for the operational choices available from varying free parameters in Figure \ref{fig:fusion_cogeneration_permutations}, as well as their impact on the risk-reward trade-off for generating revenues when electricity prices are relatively volatile compared with heat prices. High DSR means offtakers are more prepared to curtail demand.}
                \label{mat:risk_reward_spread}
        \end{figure}
        
        Figure \ref{fig:net_revenue_NPV_price_ratio} also shows the associated project NPV for the baseline case operating in each mode. We set conversion and storage CAPEX to zero for comparison. In this case, only revenues distinguish each mode, hence NPV and revenue have the same shape; additional costs introduced in the NPV calculation have no dependence on commodity prices. Notably, for the baseline case, selling heat is profitable over all price points covered. Therefore, if electricity prices were volatile, then no curtailment would be required. This is not true for selling electricity, which becomes unprofitable below the breakeven price of electricity. Of course, if heat prices were to fall then selling heat would also become unprofitable in the same way; in this case, decreasing the heat price would move all the lines representing heat output downward until the revenues generated by all modes converge on the right-hand-side of the figure – corresponding to electricity prices tending to zero.

    \subsection{\label{subsec:permissable_reductions_in_performance}Permissable reductions in performance}

\begin{figure*}[!htbp]
            \centering
                \includegraphics[width=0.8\textwidth]{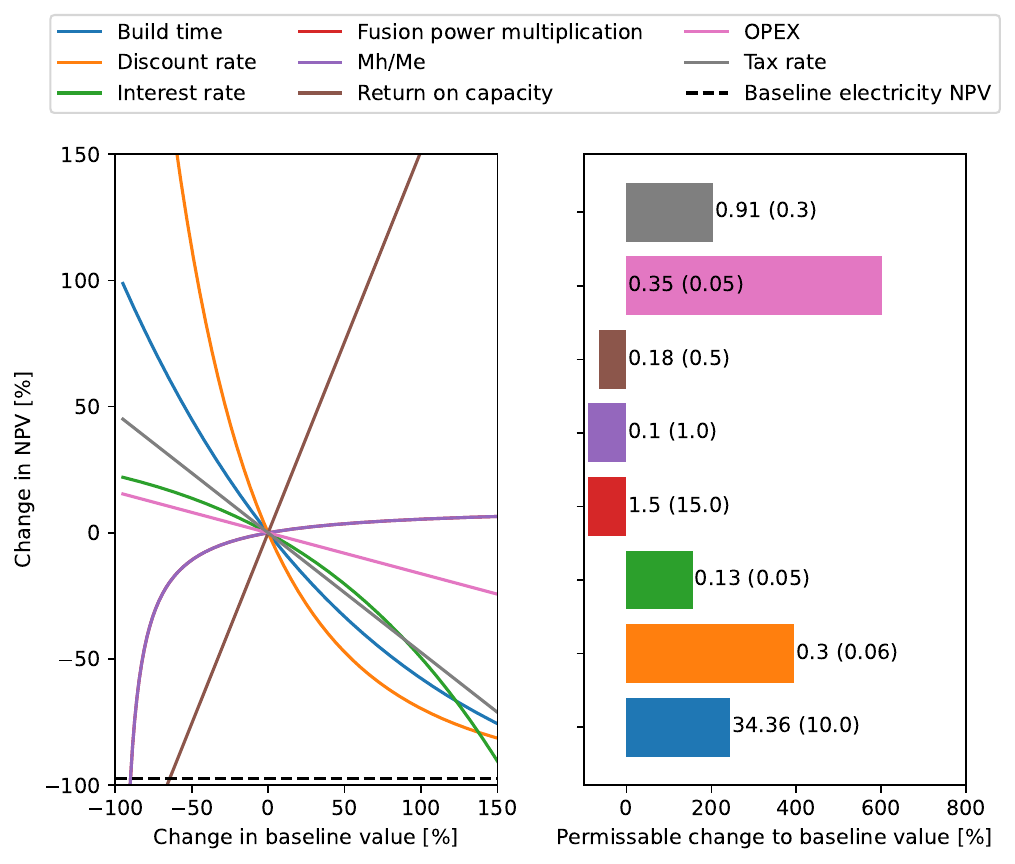}
                \caption{Left: Sensitivity analysis quantifying how changes in input parameters impact NPV. For comparison, the NPV generated by the baseline electricity plant is shown in black. Right: The change in individual parameters required to reduce NPV to that generated by the baseline electricity plant - described by the values in brackets. These changes can be interpreted as the permissable reduction in performance of a given aspect of fusion reactors if heat can be supplied. For example, the discount rate of 30\% essentially defines the Internal Rate of Return (IRR) of the heat plant. OPEX is expressed in units of overnight capital cost.}
                \label{fig:NPV_sensitivity_permissable_change}
        \end{figure*}

        The increase in profitability from selling heat can offset reductions in value elsewhere - for example decreasing $\eta_{\mathrm{fp}}$, which equates to decreasing $Q_{\mathrm{eng}}$. Figure \ref{fig:NPV_sensitivity_permissable_change} quantifies these permissable reductions: the first subplot shows the sensitivity of NPV to changes in individual parameters for the baseline heat plant, and the second subplot shows the change required in each parameter to reduce the NPV to that achieved when selling electricity - which is shown in black dash. 
        
    \subsection{\label{subsec:profit_uncertainties}Impact of model uncertainties}

\begin{figure*}[!htbp]
            \centering
                \includegraphics[width=0.7\textwidth]{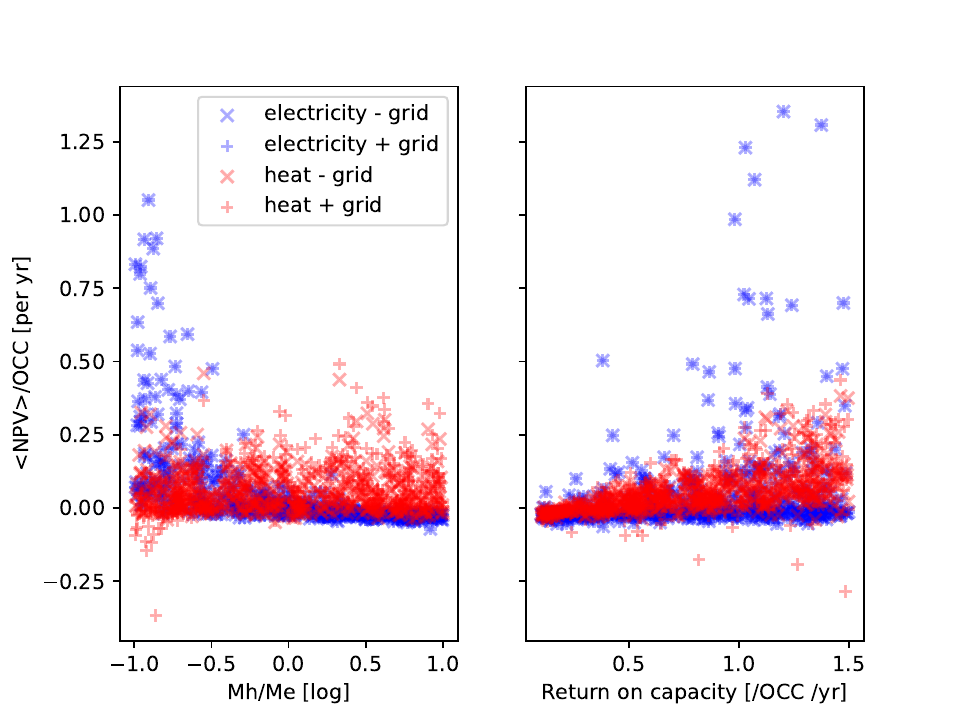}
                \caption{Average NPV generated per year as a fraction of overnight capital cost plotted against heat:electricity price ratio (left) and return on capacity (right) for a random sample of plants generated from parameter ranges given in Table \ref{tab:model_typical_parameters}. Electricity points overlie as revenues are equal in both cases. In the left figure, the moving average should approximate the NPV plotted in Figure \ref{fig:net_revenue_NPV_price_ratio}, however the spread due to uncertainties is clear. It can be seen the amount of spread is not enough to allow any significant number of electricity plants to generate profit even at price parity. In the figure on the right, the role of return on capacity as defining the envelope for possible revenues is clear; points are scattered between near-zero and a linear upper limit. Increasing OCC by e.g. 10\% will shift points leftward by 0.1.}
                \label{fig:total_NPV_monte_carlo}
        \end{figure*}

        To illustrate simply how parameter ranges propagate to measurements of the NPV, a uniform Monte Carlo sample has been generated from the ranges given in Table \ref{tab:model_typical_parameters} for each of the modes. These are plotted in figure \ref{fig:total_NPV_monte_carlo} as the average annual NPV generation. The OCC of energy storage systems for pulsed power systems was not included, since this feature is not characteristic of heat or electricity systems and therefore only serves to scale the return on capacity – shifting all the points in the right subplot of Figure \ref{fig:total_NPV_monte_carlo} below. Therefore, it could be assumed that all points here represent effectively steady-state systems. Electricity points overlay as they have equal revenues, whilst heat points are distinguished by whether they take inputs from the grid or not.

        The distributions of points against $M_{\mathrm{h}}/M_{\mathrm{e}}$ in Figure \ref{fig:total_NPV_monte_carlo} follow the revenues of the baseline plant. The range is sufficiently large that even low-risk permutations like the standalone heat mode still have some chance to be unprofitable. Selling electricity has the largest upside, but prices must be high for this to be realised; at price parity there are few ways for selling electricity to be profitable.

        Return on capacity forms the bounding envelope, within which points are spread down to a lower limit near NPV=0. Even from a relatively low return on capacity, there is the potential for significant upside when selling electricity, albeit entirely dependent on there being consistently high electricity prices. Meanwhile heat profits rise much more slowly. In both cases, there is a point where reducing the return on capacity will send average profits negative. Therefore, any headstart in increasing the return on capacity is vital. One way to achieve this is via reductions in the capacity cost, which is demonstrated here by the increase in OCC for conversion equipment required by electricity plants, shifting the blue points leftward.

\section{\label{sec:heat_or_electricity}Heat or electricity? Decarbonising heat and future price dynamics}

    Whilst so far we have estimated the potential gain in profitability from producing different outputs in the case where prices are fixed, in reality, these prices are uncertain and could vary in time. In this case, the decision to build a heat plant, an electricity plant, or a cogeneration plant depends on the potential added value from producing one output over another and the ability to dynamically switch outputs, given the current and expected future price changes \cite{hampe2016economic}. Cogeneration becomes less attractive with increasing switching costs and price stability. And in the case where heat and electricity prices become coupled, even when one is volatile, then plants that produce static outputs are always preferable – in which case condition \ref{eq:cases_operating_modes} above becomes an investment criterion, dictating whether one should construct a plant that sells heat or one that sells electricity. Therefore, we will now explore the conditions under which such price dynamics may arise in the future, and compare this to the different fusion output configurations explored above – standalone heat, the heat pump, standalone electricity, and the electricity multiplier. This requires estimating how heat prices will be formed, whilst acknowledging that electricity prices will remain highly uncertain. We start by discussing technology pathways to satisfying future heat demand, as well as how the features of these technologies will dictate the price at which competitors can produce and supply heat. We assume decarbonisation is the ultimate goal.

    There are two ways to decarbonise heat demand: with electricity or heat directly. Here we will briefly outline each.

\subsection{\label{subsec:decarbonisation_pathway_primary_heat}Decarbonisation with primary heat: standalone fusion heat competes on price and feasibility}

    Most heat demands today are met with primary heat generated by combustion of extracted fossil fuels. Either carbon capture \& storage (CCS) CAPEX will be required to maintain this or switching the heat source for an alternate technology will be required (direct-air capture (DAC) technologies are likely to be much more expensive than CCS per tonne of carbon dioxide removed). Alternatives sources of primary heat include nuclear fission, modern bioenergy, concentrated solar power (CSP), or geothermal technologies. Fusion heat fits into this category of suppliers. Heat cannot be transmitted over long distances efficiently, meaning these technologies must generate power at the point of consumption. This incentivises a monolithic approach to energy system design, with consumers co-locating with supply, joined directly not via the grid (“behind the meter”). However, all of these technologies require some electricity to operate. This can be obtained either from the grid or from the heat they produce, but for now, we will consider the latter case and ignore the supply of electricity. This corresponds to the fusion standalone heat option above. In this case, with the exception of fossil with CCS, heat produced by these technologies would be relatively stable in cost due to them having low or negligible variable costs and those costs having little exposure to market fluctuations. For example, the only volatile variable cost component of a fission plant which supplies its own electricity demand is the fuel (which is inexpensive). Therefore, with consumers not participating in markets and potentially financing their own generation assets, the heat price is formed by the total levelised cost of heat of the cheapest viable competitor technologies. Whilst it is hard to comment on how these costs may compare to fusion, especially in the future, some conclusions can be drawn with qualitative comparisons. These are summarised in Table \ref{tab:heat_competition} below, as well as for electrification, which will now be discussed.

    \begin{table*}[!htbp]
        \begin{tabular}{llllllll}
            \\ \textbf{Technologies}   & \textbf{Temperature*}                                       
            \\ Bioenergy               & \cite{silva2021analysis}                                    
            \\ CSP                     & \cite{sedighi2021experimentally}\cite{lipinski2021progress} 
            \\ Fission                 & Similar coolant principles \& technologies                             
            \\ Fossil + CCS            & \cellcolor[HTML]{FFCCC9}Widespread use today                
            \\ Geothermal              & \cellcolor[HTML]{9AFF99}\cite{jolie2021geological}          
            \\ Electrification         & \cellcolor[HTML]{FFCCC9}\cite{madeddu2020co2}               
            
            \\                         & \textbf{Siting location constraints}                                                             
            \\ Bioenergy               & \cellcolor[HTML]{FFCCC9}Fuel transportable, simple site requirements                             
            \\ CSP                     & \cellcolor[HTML]{9AFF99}Geographic/terrain constraints \cite{spyridonidou2023systematic}         
            \\ Fission                 & Similar site requirements**                                                                     
            \\ Fossil + CCS            & \cellcolor[HTML]{FFCCC9}Fuel transportable, simple site requirements                             
            \\ Geothermal              & \cellcolor[HTML]{9AFF99}Geography-dependent \cite{aghahosseini2020hot}\cite{jolie2021geological} 
            \\ Electrification         & High for indirect \cite{iea2019futurehydrogen}\cite{guerra2019cost}, less for direct             
                                    
            \\                         & \textbf{Distance between heat generation and consumption}                                                                
            \\ Bioenergy               & \cellcolor[HTML]{FFCCC9}Fuel transportable to consumption point                                                          
            \\ CSP                     & Similar heat transmission technologies                                                                                   
            \\ Fission                 & \cellcolor[HTML]{9AFF99}Licensing/regulation complicate integration \cite{angulo2012europairs}                           
            \\ Fossil + CCS            & \cellcolor[HTML]{FFCCC9}Fuel transportable to consumption point                                                          
            \\ Geothermal              & Transmission required between heat source and consumer                                                                   
            \\ Electrification         & \cellcolor[HTML]{FFCCC9}Electricity transmittable, dependent on conversion technology 
            
            \\                         & \textbf{Site power density}                                                                                    
            \\ Bioenergy               & Plant site size similar to that of fusion***                                                                    
            \\ CSP                     & \cellcolor[HTML]{9AFF99}\cite{noland2022spatial}                                                               
            \\ Fission                 & Dependent on technology, but plant output and footprint are similar                                            
            \\ Fossil + CCS            & \cite{noland2022spatial}                                                                                       
            \\ Geothermal              & \cellcolor[HTML]{9AFF99}\cite{noland2022spatial}                                                               
            \\ Electrification         & \cellcolor[HTML]{FFCCC9}Dependent on conversion technology but generation can be offsite 
            
            \\                         & \textbf{Firmness}                                                                
            \\ Bioenergy               & Both technologies firm/dispatchable                                             
            \\ CSP                     & \cellcolor[HTML]{9AFF99}Diurnal without storage                                  
            \\ Fission                 & Both technologies firm/dispatchable 
            \\ Fossil + CCS            & Both technologies firm/dispatchable                                             
            \\ Geothermal              & \cellcolor[HTML]{FFCCC9}Resources always present                                                         
            \\ Electrification         & \cellcolor[HTML]{9AFF99}Uncertain, demand-side flexibility may be required                         
            
            \\                         & \textbf{Supply security}                                                                                            
            \\ Bioenergy               & \cellcolor[HTML]{9AFF99}Low fuel energy density extends supply chain \& market exposure                                                           
            \\ CSP                     & \cellcolor[HTML]{FFCCC9}No fuel supply chain                                                       
            \\ Fission                 & Fuel energy density enables significant stockpiling but processing introduces dependencies    
            \\ Fossil + CCS            & \cellcolor[HTML]{9AFF99}Low fuel energy density extends supply chain \& market exposure, short-term
            \\ Geothermal              & \cellcolor[HTML]{FFCCC9}No fuel supply chain though localised
            \\ Electrification         & \cellcolor[HTML]{9AFF99}Long-distance interconnections introduce dependencies                                                                       
            
            \\                         & \textbf{Integrability}                                                                  
            \\ Bioenergy               & \cellcolor[HTML]{FFCCC9}Already in widespread use                                       
            \\ CSP                     & Similar technologies and requirements for heat transmission                                                
            \\ Fission                 & Similar technologies and requirements for heat transmission                                                
            \\ Fossil + CCS            & \cellcolor[HTML]{FFCCC9}Already in widespread use                                       
            \\ Geothermal              & Heat transmission also required between source and consumption                                                
            \\ Electrification         & \cellcolor[HTML]{FFCCC9}Many technologies exist today and in R\&D \cite{madeddu2020co2} 
            
            \\ & * assuming 1000C fusion
            \\ & ** technology-dependent  
            \\ & *** ignoring land use, otherwise sparse       
                
        \end{tabular}
        \caption{Table comparing potential heat supply technologies against fusion across various qualitative characteristics. Because quantitative comparisons are complex and situation-dependent, technologies are instead judged as to whether they are likely to perform better/worse for a given metric, based on robust characteristics - such as inherent or physical characteristics and current regulation. Individual reasoning is given.}
        \label{tab:heat_competition}
    \end{table*}


\subsection{\label{subsec:decarbonisation_pathway_electrification}Decarbonisation with electrification: fusion heat pump could be best}

\begin{figure}[!htbp]
        \centering
            \includegraphics[width=0.4\textwidth]{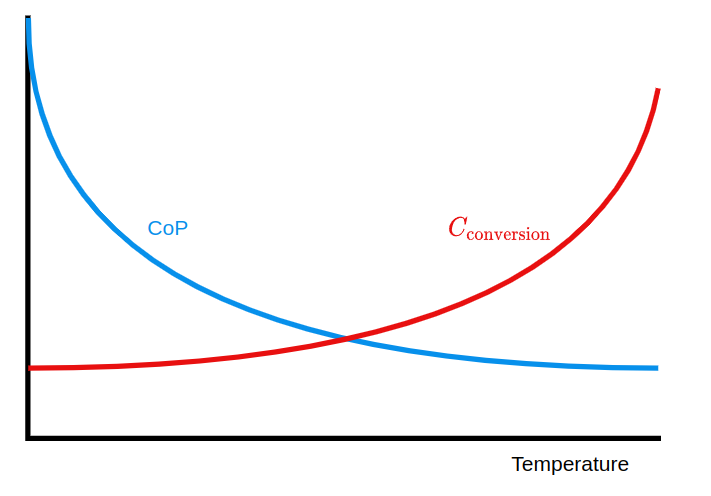}
            \caption{Illustration of general trends for coefficient of performance (COP) and conversion cost as output heat temperature increases. In both cases, higher temperatures result in higher implied costs, with COP dropping below 1 at some point for technologies today over temperatures required.}
            \label{fig:CoP_conversion_cost_temperature}
    \end{figure}

    Electrification forms a significant component of the strategy to decarbonise future global heat demand \cite{IPCC_SR15}. Direct electrification involves generating heat using conversion equipment that consumes clean electricity, likely from the grid – for example, electric arc furnaces. Meanwhile, some processes may require indirect electrification. This is achieved through an intermediate product such as hydrogen, which is synthesised with electricity – in this example with electrolysers. Electrification is a credible pathway: Some estimate that, in some regions, 78\% of heat demand could already be electrified using technology available today, with 99\% reached using technologies in development \cite{madeddu2020co2}. Hence the outstanding question is cost: of the input electricity and the additional CAPEX required to convert to heat at a given temperature. This CAPEX cost rises monotonically with the temperature requirement, as high-temperature heat can be diluted to lower temperatures at negligible cost whilst the reverse is not true (for example, coal can be equally used to boil water for cooking as it can for smelting steel in a blast furnace). This also means the relative size of the electricity component of the electrified heat cost decreases with temperature; even if the electricity price drops to zero, heat costs are finite and still increase with temperature. Therefore, in a highly electrified system, the effective heat price is coupled to the price of electricity, which remains uncertain. Which price is higher depends on the cost and efficiency (or coefficient of performance, COP) of the conversion equipment: if COP > 1 then the heat price could be lower, but if COP < 1 the heat price is always higher. Hence CoP quantifies exposure to electricity market prices; as CoP $\rightarrow \infty$, the required input electricity reaches zero and costs become dominated by the conversion equipment. COP itself is a decreasing monotonic function of temperature (a new technology with a higher CoP at a given temperature can produce heat at the same COP for all lower temperatures via dilution), meaning that there will be a temperature at which the price for heat is always greater than for electricity. These concepts are summarised in Figure \ref{fig:CoP_conversion_cost_temperature}.

\begin{figure*}[!htbp]
        \centering
            \includegraphics[width=0.7\textwidth]{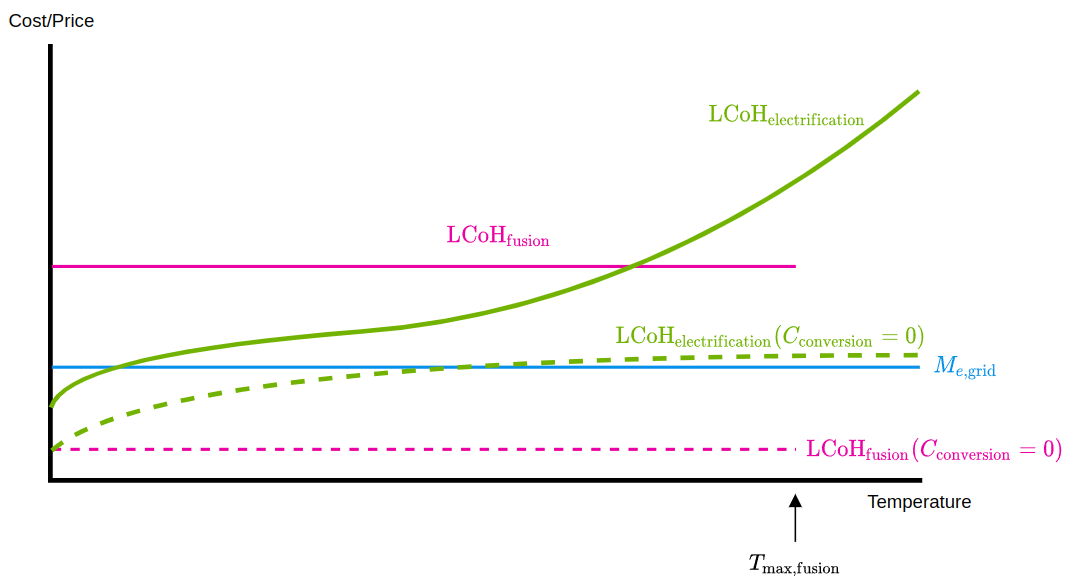}
            \caption{Heat supply curve, illustrating marginal cost of heat production for fusion and electrified systems, based on the logic from Figure 10. Curves with and without conversion costs are included, based solely on COP to illustrate where it could be possible to observe lower heat prices compared with electricity. Competing ultimately requires maximising COP, to reduce the electricity cost component of the final total heat cost, as well as minimising the remaining conversion costs that arise due to other costs, such as CAPEX for heaters or transformers. That competition occurs up to the maximum temperature at which fusion can supply heat, shown in black. }
            \label{fig:macro_heat_supply_separate}
    \end{figure*}

\begin{figure*}[!htbp]
        \centering
            \includegraphics[width=0.7\textwidth]{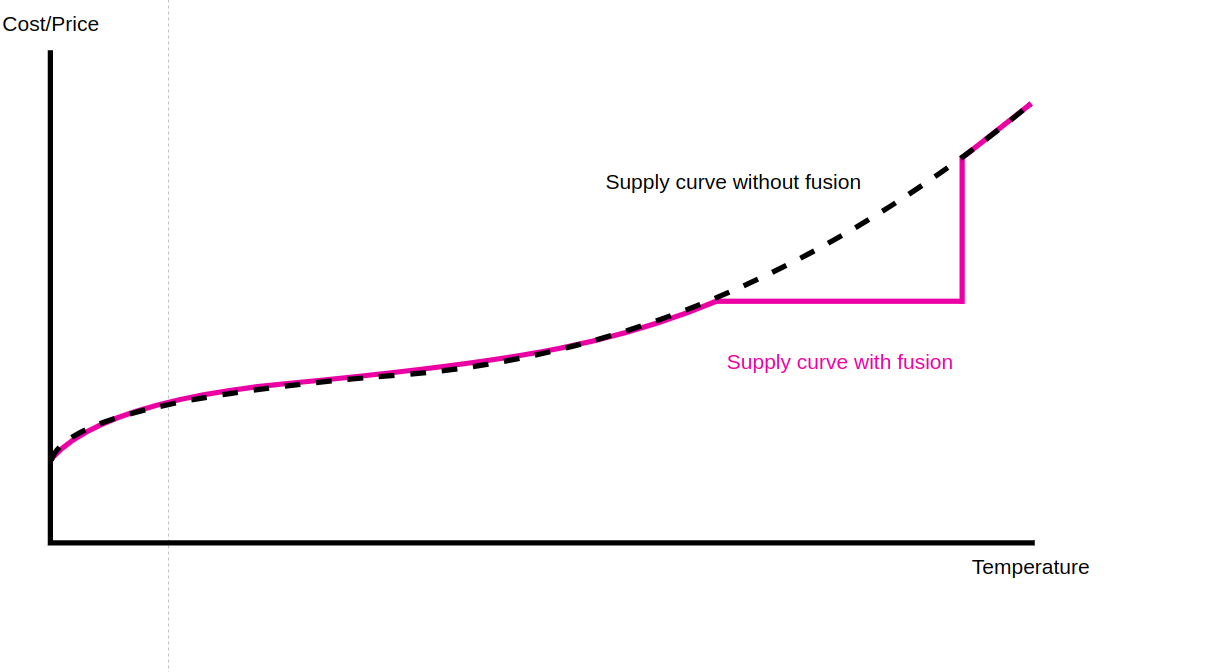}
            \caption{Aggregate supply curve for electrified heat derived from Figure \ref{fig:macro_heat_supply_separate}, with and without contributions from fusion. Fusion acts to flatten the supply curve where it can supply heat at lower cost, up to the temperature it can supply. As well as maximum temperature, this range is determined by the efficiency and cost of converting grid electricity to output heat - both for fusion and competitors.}
            \label{fig:macro_heat_supply_curve}
    \end{figure*}

    In this pathway, a fusion power plant could either supply clean electricity, which is fed into conversion equipment (standalone electricity mode or electricity multiplier mode), or it could be used as the conversion equipment itself (fusion heat pump). The first case is only preferable to directly supplying heat with fusion if electrified equipment is required to boost the output temperature, in which case the only price to consider is the uncertain market electricity price. Hence fusion would then compete on the electricity market and heat is just another demand source. The remaining case is where fusion acts as a grid-scale heat pump, multiplying grid electricity input and producing heat at a certain temperature. In this case, then the criterion for choosing to sell heat is nearly guaranteed since $\eta_{\mathrm{gen}}$ < 1 by thermodynamics and $M_{\mathrm{h}}/M_{\mathrm{e}}$ > 1 if the temperature is sufficiently high (assuming CoP falls below 1 at some point or conversion costs grow monotonically). Hence, if the grid is electrified then fusion plants should always prefer to sell heat at high temperatures. Another crucial conclusion here is that a fusion plant acting as a grid-scale heat pump effectively reaches COP > 1 at temperatures much higher than other electrified conversion technologies. Hence at high temperatures, heat from low-cost fusion plants would be cheaper than electricity prices. In this case, fusion would be guaranteed to be cheaper than competitors. That is to say, in a highly electrified system, low-cost fusion heat pumps supplying sufficiently high temperatures would always lower system costs, regardless of movements in electricity prices. This concept is illustrated in Figure \ref{fig:macro_heat_supply_separate}, where the levelised cost of heat from fusion and electrified sources is shown, in the case where conversion costs (all costs except the cost of electricity) are removed and included. Figure \ref{fig:macro_heat_supply_curve} shows the corresponding supply curve, with and without fusion included. In the case where a variety of different fusion blankets are developed, the aggregate fusion supply will also show an upward curve with temperature. 

\begin{figure}[!htbp]
        \centering
            \includegraphics[width=0.3\textwidth]{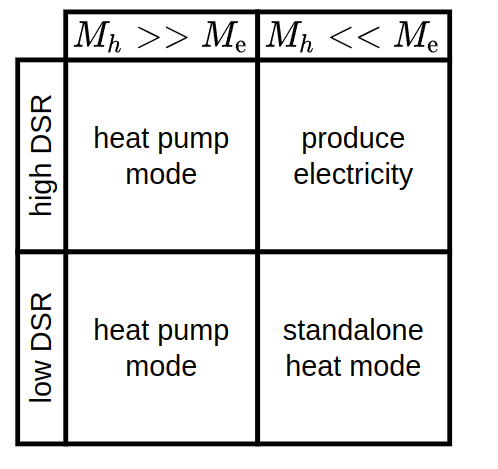}
            \caption{Decision matrix for operating strategy given criteria for demand-side response flexibility and heat:electricity price ratio trends. "Produce electricity" refers to either strategy - with or without grid input. Where DSR can be tolerated, plants are encouraged to switch output types based on market fluctuations, whilst low DSR tolerance requires plants only to decouple from the grid to mitigate high electricity prices, at the cost of reducing output due to recirculation requirements. Therefore, in any case where fusion plants are coupled to volatile electricity markets, some risk will need to be tolerated by plant managers or offtakers.}
            \label{mat:DSR_price_ratio}
    \end{figure}

    Combining the results so far, considering the trade-offs of selling heat and electricity whilst acknowledging volatility in the electricity market, there may be potential for a second class of plants: ones which combine the characteristics above by possessing the ability to switch input and output streams flexibly whilst operating continuously. Cogeneration plants are one such example, but there are more permutations. Considering the heat plant operating in heat pump mode primarily, it may become optimal to purchase the option to operate in standalone heat mode to hedge against sudden large increases in electricity prices. Given the required reduction in heat output, some demand-side response (DSR) must be permitted – likely at additional cost. If demand flexibility is increased even further, likely at greater cost, it could become more optimal to switch output completely to supplying electricity. This strategy is shown in Figure \ref{mat:DSR_price_ratio}. The technical requirements and costs for achieving this level of flexibility warrant further examination.

\section{\label{sec:discussion}Discussion}

    This work illustrates the degree to which selling heat could facilitate the commercialisation of fusion. The increase in profitability and favourable competitive dynamics could offer sufficient prize for crossing the valley of death \cite{cardozo2019economic} and research parallelisation \cite{cardozo2024interplay}, bootstrapping learning, and driving costs down to make grid electricity more approachable. Prospects could be improved further if heat generation results in new design criteria that lead to simplification and optimisation - for example, removing conversion and storage CAPEX, capturing waste streams, and re-designing pipework. These benefits may be translatable to the electricity-producing plants under development, as has been identified by Cano-Megias et al. \cite{cano2022boosting} when discussing fusion cogeneration with heat waste streams. However, care should be taken to distinguish “profitable” and “competitive” when valuing these potential gains. Fusion heat must be both.

    Key to this analysis are heat transmission costs, which so far have been assumed to equal those for electricity. This is unlikely to be the case for high-temperature heat, in which case providing industrial heat directly with a fusion plant may have important consequences for energy system design. The high value density of fusion power, combined with physical transmission constraints, means offtakers must site themselves as close as possible to singular fusion power sources. Tightly-coupled industrial systems, based around centralised, monolithic power sources already exist in the form of eco-industrial parks like that in Kalundborg in Denmark, and the Maasvlakte in the Netherlands. Similarly, large-scale hydrogen infrastructure is most viable if arranged in similar clusters \cite{iea2019futurehydrogen}. Likewise, such monolithic systems are vital for building secure energy systems in high-density geographies, such as Singapore and Taiwan.

    In such a configuration, new lines of competition will also emerge. For example, siting and proximity are potential advantages fusion possesses over fission, and potentially other forms of direct heat. Compared with electrified heating, especially low-CoP technologies, captive fusion plants could offer increased certainty to offtakers by decoupling heat costs from volatile electricity prices - forming a heat-as-a-service model. At the larger scale, the choice between supplying heat via electrification or directly introduces competition over design philosophy: that of the monolithic system of distributed energy resources (DER) versus grid-based. Competitions between these analogous architectures have appeared elsewhere, such as the move toward extensible, standardised supercomputing architectures. 

    \subsection{\label{subsec:future_work}Future Work}

        The model used herein contains some significant sources of uncertainty. Whilst heat and electricity transmission efficiencies are assumed equal, in reality they are not – which would impact the conclusions of this study, as well as having implications for similar competing technologies, such as fission and CSP. Answering this question - the third outlined in the Introduction – is of crucial interest if fusion is to leverage the significant commercial opportunity offered by heat. However, research on this topic – particularly in the fusion domain – is rare.
    
        Another source of uncertainty is in the neglected capacity factors. Whilst operational experience will be the only true remedy, it is valuable to theoretically understand how capacity factors may be improved by cogeneration, where flexible demand could mitigate risks. Both of these factors - transmission efficiency and capacity factor - directly affect revenues, making them important parameters. Further studies should aim to realistically evaluate these parameters as anecdotal evidence appears or bottom-up analysis is enabled.
        
        Finally, whilst physics principles have been applied to increase the robustness of the economic conclusions herein, significant uncertainties still remain that will distort these findings. Heat technology, power load, and process type introduce context dependency in the formulation of heat supply curves. Measuring this distortion is crucial for evaluating the competitive prospects of fusion in local contexts; heat is not, and never likely to be, traded on a distributed and deregulated marketplace, which makes the specific use context important. And with only a handful of specific end-uses comprising a significant share of demand today, more detailed studies of fusion supplying heat to specific processes is warranted. Nor can these insights be gained elsewhere; little literature exists on assembling supply curves for heat, as heat is dominated by fossil today, which can cheaply supply heat at most temperatures. 

    \subsection{\label{subsec:implications_strategy_policy}Implications for strategy and policy}

        Some strategic actions are now outlined.
            
        \begin{itemize}    
            \item \textit{For tightly-coupled energy systems, avoiding lock-in \cite{janipour2022industrial} and achieving harmony amongst components requires facilitating organic system growth \cite{perrucci2022review}. Smaller, standardised and modular units could be preferable \cite{carelli2010economic}\cite{boarin2021economics}. Standardisation is especially true for fusion device blankets. As the interface between the fusion core and specialised commercial output \cite{vanatta2023technoeconomic}, a standardised blanket could act as a design pivot point, around which fusion core designs could vary whilst maintaining commercial integration and providing a platform for consumers to adapt around.} Such a technology platform \cite{ansar2022solve} could become a tangible output from public-private partnerships that are now gaining traction.
            \item \textit{Technologies for the high-efficiency transmission and exchange of high-temperature heat should be considered critical to the fusion commercialisation mission. Solutions to the associated materials challenges \cite{ibano2011design} are likely to spillover to fission \cite{energyiaea}, CSP technologies, and beyond.}
            \item \textit{Strategies for addressing the regulatory challenges of tightly integrating fusion with industry should be developed. }
            \item \textit{Seriously consider the option value of conversion equipment installation in the case where fusion supplies heat. Switching input sources to self-generated electricity hedges curtailment risk during electricity price spikes, increases portability, and increases availability in the case where the other input channel fails. Switching output sources could enable fusion to displace renewable overcapacity and low-utilisation grid-scale batteries in the case where renewable supply severely drops; in such rare cases, fusion heat can be replaced with alternative dispatchable heat sources (such as fossil with CCS) at a higher efficiency than alternatively using excess renewable electricity stored over long periods, which would require significant overcapacity.}
        \end{itemize}

\section{\label{sec:summary}Summary}

    This paper explores the economic potential of fusion reactors which aim to supply heat, with or without electricity – a topic that is mostly missing from fusion research. A simple, generalised model was used to quantify the opportunity that heat presents for fusion profitability and adoption readiness levels.

    For plants described in literature, the decision to sell heat or electricity could be the factor deciding whether those plants are profitable or not. Unlike other fusion studies, which typically focus on cost of production alone, this considers the market context, where commodity prices trade near electricity prices today. By examining the potential structure of these markets, as well as fusion’s competitors and their technological characteristics, from a qualitative and physics point of view, broad conclusions can be drawn on determinants for success of fusion heat: temperature, cost, and system design. If fusion were to be deployed in the medium term, it would likely compete against other potential captive plants that supply direct, primary heat. Low costs \cite{gilbert2023heat} makes fossil with carbon capture the likeliest competitor. Alternatively, changes to the competitive landscape from long-term electrification may require fusion to consider alternative roles that require operational flexibility. 

\section{\label{sec:acknowledgements}Acknowledgements}

    \textit{This work has been carried out within the framework of the EUROfusion Consortium, funded by the European Union via the Euratom Research and Training Programme (Grant Agreement No 101052200 - EUROfusion). Views and opinions expressed are however those of the author(s) only and do not necessarily reflect those of the European Union or the European Commission. Neither the European Union nor the European Commission can be held responsible for them.}

\section*{\label{sec:references}References}
    \scriptsize 
    \bibliographystyle{unsrt}
    \bibliography{references}
\end{document}